# Atomically Thin Boron Nitride as an Ideal Spacer for Metal-Enhanced Fluorescence

Wei Gan,[1] Christos Tserkezis,[2] Qiran Cai,[1] Alexey Falin,[1] Srikanth Mateti,[1] Minh Nguyen,[3] Igor Aharonovich,[3] Kenji Watanabe,[4] Takashi Taniguchi,[4] Fumin Huang,[5] Li Song,[6] Lingxue Kong,[1] Ying Chen[1] and Lu Hua Li[1]*

[1] Institute for Frontier Materials, Deakin University, Geelong Waurn Ponds Campus, Victoria 3216, Australia

[2] Center for Nano Optics, University of Southern Denmark, Campusvej 55, DK-5230 Odense M, Denmark

[3] School of Mathematical and Physical Sciences, University of Technology Sydney, Ultimo, New South Wales 2007, Australia

[4] National Institute for Materials Science, Namiki 1-1, Tsukuba, Ibaraki 305-0044, Japan

[5] School of Mathematics and Physics, Queen's University Belfast, Belfast BT7 1NN, United Kingdom

[6] National Synchrotron Radiation Laboratory, CAS Center for Excellence in Nanoscience, University of Science and Technology of China, Hefei, Anhui 230029, China

* E-mail: luhua.li@deakin.edu.au

**ABSTRACT**

The metal-enhanced fluorescence (MEF) considerably enhances the luminescence for various applications, but its performance largely depends on the dielectric spacer between the fluorophore and plasmonic system. It is still challenging to produce a defect-free spacer having an optimized thickness with a subnanometer accuracy that enables reusability without affecting the enhancement. In this study, we demonstrate the use of atomically thin hexagonal boron nitride (BN) as an ideal MEF spacer owing to its multifold advantages over the traditional dielectric thin films. With rhodamine 6G as a representative fluorophore, it largely improves the enhancement factor (up to $\sim 95 \pm 5$), sensitivity ($10^{-8}$ M), reproducibility, and reusability ($\sim$90% of the plasmonic activity is retained after 30 cycles of heating at 350 °C in air) of MEF. This can be attributed to its two-dimensional structure, thickness control at the atomic level, defect-free quality, high affinities to aromatic fluorophores, good thermal stability, and excellent impermeability. The atomically thin BN spacers could increase the use of MEF in different fields and industries.

**KEYWORDS:** boron nitride; metal-enhanced fluorescence; dielectric spacer; plasmonic nanoparticle; fluorescence quenching; two-dimensional materials

Fluorescence could be very useful in chemical detection, biomedical imaging/diagnosis, mineralogy, optoelectronics, and forensics.[1-4] However, its small intensity hinders wider applications. The luminescence intensity can be considerably amplified by metal-enhanced fluorescence (MEF), which uses localized surface plasmon resonance (LSPR) to increase the light absorption and emission cross sections of fluorophores.[5-7] However, quenching occurs when fluorophores are in close proximity to metals owing to the transfer of the excited electrons to the nearby metals *via* nonradiative decay instead of emission.[8] Therefore, a dielectric spacer needs to be inserted between the fluorophore and plasmonic metal for a minimized quenching and maximized luminescence. Both plasmonic enhancement and fluorescence quenching are extremely sensitive to distance; *e.g.*, the plasmon effects diminish exponentially with the increase in distance. Therefore, the spacer thickness should be precisely controlled at the atomic level. In addition, the spacer needs to have a very high quality to avoid defect-caused tunneling and quenching.[9] The dielectric properties of the spacer also affect the plasmonic enhancement. Another practical function of the spacer is to protect air-sensitive metal nanoparticles (NPs). Ag NPs under ambient conditions lose 26% and 52% of the plasmonic enhancements owing to oxidation after 1 and 5 days, respectively.[10] We observed reductions in plasmonic activity of Ag NPs of 32% and 81% after heatings at 360 °C in air for 5 and 35 min, respectively.[11] Therefore, the ideal spacer should protect the plasmonic NPs during heating in oxygen-containing environments to burn off the organic fluorophores to achieve MEF reusability.

Polymers[12,13] and ceramics, such as alumina ($Al_2O_3$)[14] and silicon oxide ($SiO_2$),[15] are traditional spacer materials. The main disadvantage of the polymers is that a relatively thick layer (20–40 nm) is required to attenuate the quenching owing to the presence of defects.[13] Furthermore, polymer films usually cannot be fabricated with atomically precise thicknesses and are not thermally robust to realize reusability. In contrast, the thickness of a ceramic coating could be tailored with an atomic accuracy, particularly by atomic layer deposition (ALD). The optimal thickness of an ALD-grown ceramic spacer is usually ~10 nm[14], considerably smaller than that of the polymer, as the ceramic film fabricated by ALD has a relatively high quality.

Nevertheless, the formations of defects are still inevitable in these films,[16] which is detrimental to the fluorescence intensity and protective effectiveness. Therefore, ~40 nm metal oxide layers are required for oxidation protection and reusability despite their excellent thermal stabilities.[10,11] The additional thickness reduces the MEF intensity and sensitivity. Another disadvantage of ceramic spacers are their low affinities to most organic fluorophores, restricting the detection limit of MEF.

Two-dimensional (2D) materials have attractive properties to work with plasmonic structures. Graphene was used as the thinnest spacer between metal NPs and mirror for a maximum plasmonic field enhancement.[17,18] In addition, it largely improved the sensitivity of surface-enhanced Raman spectroscopy (SERS).[10,19,21] However, the electric conductivity makes graphene unsuitable as a MEF spacer. Atomically thin hexagonal boron nitride (BN) is an electrically insulating member of the 2D material family, whose bandgap is almost independent of the thickness reduction.[22] Due to its layered structure, the thickness of the BN spacer can be well controlled at the subnanometer level (*i.e.*, 0.334 nm for each layer of BN). Monolayer BN can sustain 800 °C in air;[23] in contrast, graphene starts to oxidize at ~300 °C under the same conditions.[24] In addition, atomically thin BN is one of the strongest materials[25] and highly impermeable to gas and ions. Therefore, atomically thin BN is an excellent candidate for metal protection,[26,27] without galvanic corrosion associated with graphene.[28,29] Metal NPs protected by atomically thin BN could sustain high temperatures to remove organic molecules.[11,30,31] Furthermore, atomically thin BN has a better surface adsorption capability than that of its bulk counterpart due to conformational changes.[32] Therefore, atomically thin BN could be an ideal spacer material for MEF. However, few related studies have been reported.[33]

In this study, we show that atomically thin BN largely improved the sensitivity, stability, reproducibility, and reusability of MEF. Ag NPs covered by high-quality defect-free BN sheets provided one of the best reported enhancement factors (EFs) (95±5) of rhodamine 6G (R6G), more than two times that of the same Ag NPs coated by a traditional $Al_2O_3$ spacer produced by ALD. Owing to the BN's higher affinities to aromatic fluorophores, the BN sheets increased the fluorescence detection sensitivity up to three times, compared to the $Al_2O_3$ spacer. However, to achieve these optimal enhancements, the thickness of BN had to be precisely tuned at the atomic level. Furthermore, the BN sheets effectively protected the underlying Ag NPs from degradation at high temperatures so that the MEF substrates could be regenerated and reused.

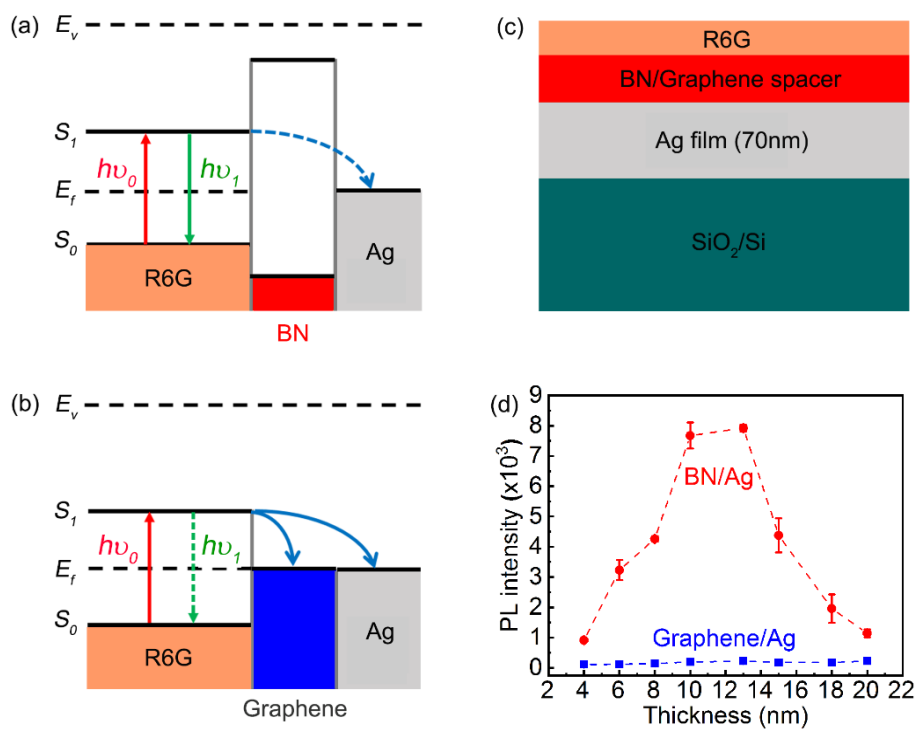

**Figure 1. Different roles of BN and graphene spacers in the MEF. (a), (b) Comparison of the energy diagrams of the R6G fluorophores on the BN/Ag and graphene/Ag substrates, where $E_v$ and $E_f$ are the vacuum energy and Fermi levels, while $S_1$ and $S_0$ are the ground and excited states of R6G, respectively; (c) schematic of the MEF structure used to**

**experimentally compare the different effects of BN and graphene as spacers; (d) PL intensities of R6G on the BN/Ag and graphene/Ag substrates with different BN and graphene thicknesses. The error bars for graphene/Ag in (d) are too small to observe.**

## RESULTS AND DISCUSSION

BN and graphene spacers have distinct effects on the MEF. Figure 1a and 1b compare the energy diagrams of R6G fluorophores and Ag separated by BN and graphene sheets, respectively. The electric insulation of BN inhibits the transfer of electrons in the excited state $S_1$ of R6G to the metal; in contrast, the excited electrons spontaneously transfer to graphene/Ag, leading to fluorescence quenching. To experimentally evaluate the difference, high-quality BN and graphene sheets having thicknesses of 4–20 nm were mechanically exfoliated on 70 nm-thick Ag films sputtered on silicon oxide-covered silicon wafers ($SiO_2$/Si) (Figure 1c). Typical atomic force microscopy (AFM) images of the samples are presented in Supporting Information (Figure S1a and S1b). R6G molecules were spin-coated on top of BN/Ag and graphene/Ag under the same conditions. As expected, the photoluminescence (PL) signal from graphene/Ag was negligible regardless of the thickness change of graphene (blue in Figure 1d). In comparison, the PL signal from BN/Ag was considerably stronger but varied with the BN thickness (red in Figure 1d). The corresponding PL spectra are presented in Supporting Information (Figure S1c and S1d). When a fluorophore is in the vicinity of a plasmonic metal, the competition between the plasmon enhancement and quenching determines the MEF performance. Two factors can contribute to the enhancement: the plasmon-enhanced excitation and antenna effect.[34,35] In the plasmon-enhanced excitation, plasmonic metal NPs enhance the local electromagnetic fields proportional to $e^{-d}$ ($d$ is the distance between the fluorophore and plasmonic system), increasing the absorption cross section of the fluorophore, following the Fermi's golden rule.[36] Regarding the antenna effect, a surface plasmon couples with the excited state of the fluorophore to escalate its radiative decay rate. The quenching rate is proportional to $d^{-4}$.[37] Consequently, the MEF intensity initially increases and then decreases with the increase in spacer thickness. Our results in Figure 1d, which follow this trend, show that the subtle thickness change of BN largely affects the MEF intensity; the optimized thickness should be between 10 and 14 nm.

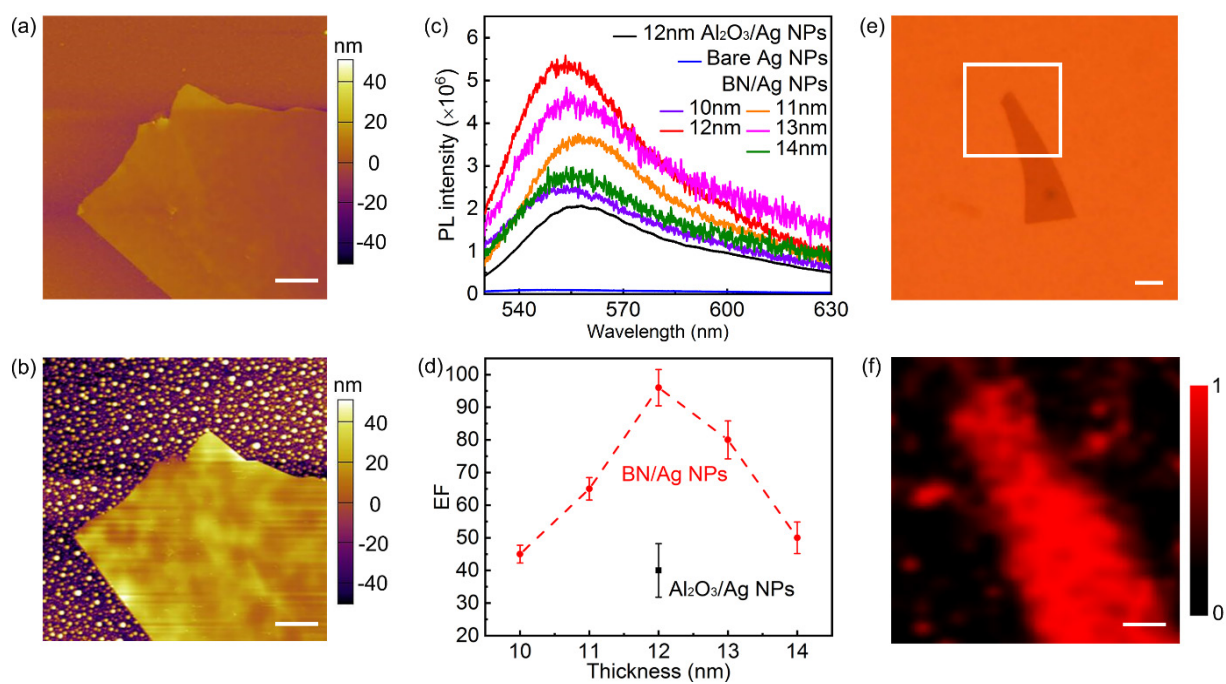

**Figure 2. MEF from the thin BN sheet-covered Ag NPs. AFM images of the 12 nm-thick BN sheet on the 12 nm Ag film on SiO₂/Si (a) before and (b) after the annealing at 400 °C (scale bar: 2 μm); (c) representative PL spectra of R6G on the Ag NPs covered by BN sheets having different thicknesses (10–14 nm) and 12 nm Al₂O₃ and that of the structure without spacer (bare Ag NPs); (d) fluorescence EFs as a function of the BN thickness; (e) optical image of the 12 nm BN-covered Ag NPs (scale bar: 5 μm); (f) corresponding**

**fluorescence map of the spin-coated R6G within the white square area shown in (e), where the color scale bar between 0.0 and 1.0 represents the weakest to the strongest fluorescence, respectively (scale bar: 2 μm).**

To maximize the fluorescence enhancement, we focused on the BN sheets having thicknesses of 10–14 nm and turned the Ag films to more plasmonically effective Ag NPs by annealing. The Ag particle size and inter-particle distance could be well controlled by varying the Ag thickness and annealing temperature.[11,30,31] As the MEF largely depends on the overlap between the LSPR spectrum and excitation spectrum of the fluorophore,[38] we optimized the LSPR band of the Ag NPs by tuning the thickness (8–16 nm) and annealing temperature (400, 450, and 500 °C) of the Ag film. The coupling was most effective after the 12 nm Ag film was annealed at 400 °C (see Supporting Information, Figure S2a–c). According to scanning electron microscopy (SEM) investigations, these Ag NPs had an average diameter of ~80 nm and inter-particle distance of ~10 nm (see Supporting Information, Figure S2d and S2e). AFM images of the 12 nm-thick BN before and after the annealing at 400 °C are shown in Figure 2a and 2b, respectively.

Figure 2c compares the PL spectra of R6G spin-coated on the 10–14 nm BN/Ag NPs. The 12nm BN provided the best enhancement with a fluorescence intensity of $5.5 \times 10^6$ and wavelength centered at 553 nm. All other BN thicknesses led to smaller fluorescence intensities at slightly different wavelengths. For comparison, we also deposited a 12 nm-thick $Al_2O_3$ on the same Ag NPs by ALD. $Al_2O_3$ is one of the most used ceramic spacers for MEF; the thickness of 12 nm is close to the optimal thickness.[14] However, its fluorescence intensity, 2.0 $\times\ 10^6$, was much lower than that of the 12 nm BN spacer. The signal of the bare Ag NPs without spacer was approximately six orders of magnitude weaker owing to the quenching effect.[8]

To quantify the enhancements of the different spacers and compare them to those in previous reports, we calculated the EF:[39]

$$EF = \frac{I_{fluorophore/spacer/Ag/SiO_2} - I_{spacer/Ag/SiO_2}}{I_{fluorophore/spacer/SiO2} - I_{spacer/SiO2}}, \qquad (1)$$

where $I_{fluorophore/spacer/Ag/SiO2}$ is the PL intensity of the fluorophore on the plasmonic Ag NPs covered by BN or $Al_2O_3$ spacer; $I_{spacer/Ag/SiO2}$ is the background PL intensity of BN/Ag/SiO$_2$ or $Al_2O_3$/Ag/SiO$_2$ without R6G; $I_{fluorophore/spacer/SiO2}$ is the PL intensity of the fluorophore on BN/SiO$_2$ or $Al_2O_3$/SiO$_2$ (without Ag NPs); and $I_{spacer/SiO2}$ is the background fluorescence intensity of BN/SiO$_2$ or $Al_2O_3$/SiO$_2$. The typical values of $I_{spacer/Ag/SiO2}$, $I_{fluorophore/spacer/SiO2}$ and $I_{spacer/SiO2}$ are presented in Supporting Information (Figure S3 and S4). The obtained EFs are plotted in Figure 2d. The largest EF of $\sim 95 \pm 5$ was calculated for the 12 nm BN/Ag NPs, more than two times that of the $Al_2O_3$/Ag NPs of $\sim 40 \pm 8$. This enhancement was one of the highest among those of previously reported MEFs of R6G. Complex fabrication processes were required to achieve similar EFs.[40,41] In addition, the PL signal from our MEF substrate was homogeneous. Figure 2e shows an optical microscopy image of the 12 nm BN/Ag NPs, while Figure 2f shows a PL map of the square area in Figure 2e. The plasmonic Ag NPs with and without BN coverage exhibited very different luminescence intensities. The weaker fluorescence from the edge of the BN sheet is attributed to the averaging of the signals from the BN/Ag NP and bare Ag NP areas.

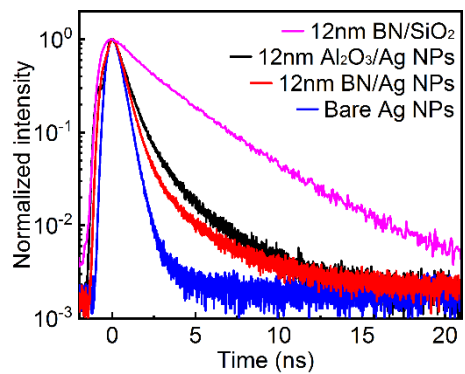

**Figure 3. Fluorescence lifetime. The lifetimes of R6G on the bare Ag NPs (0.46 ns), 12 nm BN/Ag NPs (1.12 ns), 12 nm Al₂O₃/Ag NPs (2.06 ns), and 12 nm BN/SiO₂ (3.83 ns) are compared.**

To elucidate the enhancement mechanism, we measured the fluorescence lifetimes of R6G on the different substrates, including the bare Ag NPs (0.46 ns), 12 nm BN/Ag NPs (1.12 ns), 12 nm Al₂O₃/Ag NPs (2.06 ns), and 12 nm BN/SiO₂ (3.83 ns) substrates (Figure 3). The Ag NPs greatly reduced the lifetime. Without the plasmon effect, the emission into the far-field can be described by the pristine or free-space quantum yield ($Q_0$) and fluorescence lifetime ($\tau_0$),[42]

$$Q_0 = \frac{k_0}{k_0 + k_{nr}}, \qquad (2)$$

$$\tau_0 = \frac{1}{k_0 + k_{nr}}, \qquad (3)$$

Where $k_0$ and $k_{nr}$ are the radiative and non-radiative decay rates, respectively.[43] When the fluorophore is in the close proximity of the plasmonic metal NPs, the quantum yield ($Q_m$) and fluorescence lifetime ($\tau_m$) change to[44,45]

$$Q_m = \frac{k_0 + k_m}{k_0 + k_m + k_{nr} + k_{mnr}}, \qquad (4)$$

$$\tau_m = \frac{1}{k_0 + k_m + k_{nr} + k_{mnr}}, \qquad (5)$$

Where $k_m$ is the additional radiative decay rate owing to the coupling between the fluorophore and plasmon;[46] $k_{mnr}$ is the additional nonradiative decay rate due to the absorption by plasmonic NPs.[47,48] These explain the observed reductions in lifetime with the presence of the Ag NPs (Figure 3). Our finite-element calculations showed that $k_{mnr}$ was nearly constant for BN/Ag NPs and $Al_2O_3$/Ag NPs; and the defects in $Al_2O_3$ should cause more quenching and hence more increase in $k_{nr}$.[49] However, the 12 nm BN/Ag NPs still showed a shorter lifetime than that of the 12 nm $Al_2O_3$/Ag NPs. It strongly suggest the BN/Ag NPs had much larger $k_m$ than that of $Al_2O_3$/Ag NPs due to the antenna effect. That is, the BN sheet enabled a stronger coupling between R6G and the plasmonic Ag NPs, which explains the more efficient increase in $Q_m$. The bare Ag NPs led to the smallest lifetime (0.46 ns) and weakest fluorescence, because when the molecules sit directly on Ag NPs quenching[37] dominates the Purcell factor, obtained as $\tau_0$ /$\tau_m$[47].

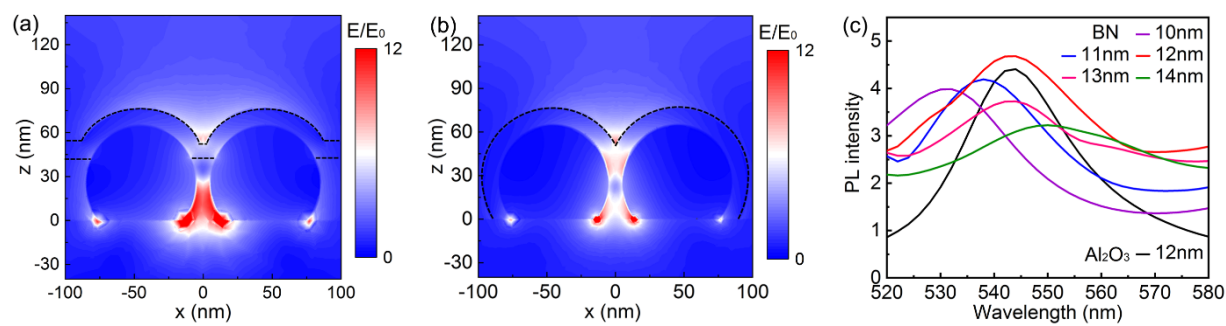

**Figure 4. Theoretical simulations. Calculated electric field enhancements in the x–z plane for the (a) 12 nm BN/Ag dimers and (b) 12 nm $Al_2O_3$/Ag dimers; (c) simulated fluorescence spectra for the different MEF substrates.**

We also performed finite-element calculations to better understand the origin of the MEF from the Ag NPs covered by the different spacers. All geometries of the models were deduced from our experiments; *e.g.,* the Ag NP height was obtained by the AFM, while the diameter and inter-particle distance were obtained by the SEM. The R6G fluorophores were modelled as point dipoles on top of the BN- and $Al_2O_3$-covered Ag NP dimers (Supporting Information, Figure S5). The plasmon-enhanced electric field was polarized along the *x* axis, which was also the plasmonic dimer axis. The near-field intensities of the 12 nm BN/Ag and 12 nm $Al_2O_3$/Ag dimer structures in the *x*–*z* plane are shown in Figure 4a and 4b, respectively. In general, the electric field on the BN spacer was stronger than that of $Al_2O_3$, which agrees well with the lifetime results that the 12 nm BN/Ag NPs had larger $k_m$ than that of $Al_2O_3$/Ag NPs. In the calculation of the fluorescence spectra, both electric field and plasmon effects were included. As mentioned above, the *x* component of the electric field was considerably stronger, which should excite dipoles with the same orientation more efficiently. Therefore, we needed to consider the dipole orientation of R6G, an elongated molecule with its dipole moment along the xanthene group. We theoretically and experimentally determined that R6G physisorbed on BN preferred the so-called lying-down orientation, in which the xanthene is parallel to the surface of BN.[32] In this regard, the R6G molecules on BN should be oriented along the *x* axis in the simulation. However, the orientation of R6G on $Al_2O_3$ has not been reported. We had to assume that the R6G molecules on top of the Ag NPs were oriented along the *x* axis, similar to the case of BN, but R6G in the valley between the two Ag NPs was oriented along the *z* axis (see Supporting Information, Figure S5). The simulated fluorescence spectra qualitatively reproduced the trend observed in our experiment that the 12 nm BN/Ag NPs had the best MEF performance and that the fluorescence wavelength slightly changed with the variation in BN thickness (Figure 4c). The different tunneling properties of the two spacer materials were not considered. Particularly, the $Al_2O_3$ thin film used in the experiment had many defects (further evidence is presented below), which facilitated the quenching. This explains the weaker fluorescence of the 12 nm $Al_2O_3$/Ag NPs in our experiment (Figure 2c). The calculations showed that both electric field and plasmon effects contributed to the large EF and that the plasmonic geometry of our Ag NPs was well suited for the fluorescence enhancement of R6G.

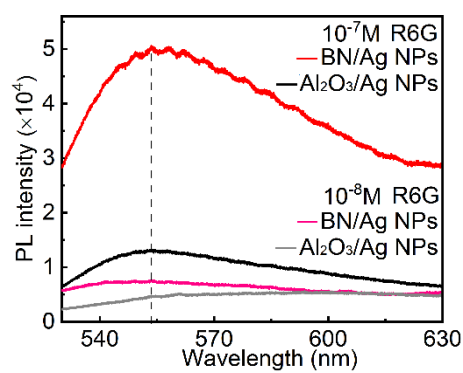

**Figure 5. Fluorescence detection limit. Fluorescence intensities of the 12 nm BN/Ag NP and 12 nm $Al_2O_3$/Ag NP substrates immersed in $10^{-7}$ and $10^{-8}$ M R6G solutions for 1 h.**

The fluorescence detection limit of the 12 nm BN/Ag NPs was evaluated and compared to that of the 12 nm $Al_2O_3$/Ag NPs. Different from the spin coating used for the EF calculations, the substrates were immersed in $10^{-7}$ and $10^{-8}$ M R6G solutions for 1 h. Although the EF cannot be quantified in this case, their difference was more striking at $10^{-7}$ M, reflected by the PL intensity ratio between the two MEF substrates, increased from $5.5 \times 10^6$ *vs.* $2.0 \times 10^6$ (*i.e.,* 2.75:1) (Figure 2c) to $5.0 \times 10^4$ *vs.* $1.3 \times 10^4$ (*i.e.,* 3.85:1) (Figure 5). This can be attributed to the higher surface affinity of BN to R6G than that of the ceramic $Al_2O_3$. Aromatic molecules and BN have π–π interactions, which can cause conformational changes in atomically thin BN, leading to increased adsorption energy and efficiency.[32] In contrast, the interactions between R6G and ceramic $Al_2O_3$ were weak van der Waals interactions. In other words, the BN physisorbed more fluorophores than $Al_2O_3$ during the immersion and hence exhibited the stronger fluorescence. The enhancement by the BN/Ag NPs was also more prominent than that by the $Al_2O_3$/Ag NPs at $10^{-8}$ M, though the signals were considerably weaker for both substrates (Figure 5).

Besides R6G, we also evaluated another fluorophore, rose bengal (RB), which has a small quantum yield. The EF of the RB on the 12 nm BN/Ag NPs was ~10.0, larger than that (~6.2) on the $Al_2O_3$/Ag NPs (Supporting Information, Figure S6–8). The considerably lower EFs of the RB than those of R6G agree with the previous study, which showed that the EF was not inversely proportional to the quantum yield.[42]

The excellent thermal stability and impermeability of the BN enable a longer-term usage of the MEF substrate and its reusability. This is beneficial particularly to the air-sensitive Ag NPs whose plasmonic efficiency decreases with the oxidation.[10,11] We evaluated the protective effectiveness of the BN spacer under harsher conditions, *i.e.*, upon heating at 350 °C in air. As the heating treatment was the most effective approach to remove organic fluorophores,[50,51] the test also demonstrated the reusability of the BN-covered MEF substrates.[11,31] In each test cycle, R6G molecules were spin-coated on the different substrates for PL measurements and then burnt off by heating in air for 5 min for regeneration. The relative changes in fluorescence intensities for the bare Ag NP, 12 nm $Al_2O_3$/Ag NP, and 12 nm BN/Ag NP substrates during the cycling are compared in Figure 6. The bare Ag NPs exhibited the lowest reusability, whose fluorescence intensity decreased by 45% and 70% after 3 and 7 cycles, respectively (black).

The 12 nm $Al_2O_3$ provided a trivial protection, *i.e.*, the fluorescence decreased by 70% after 10 cycles. This suggests the presence of many defects in the ALD-grown $Al_2O_3$ film, causing an additional quenching not included in the theoretical simulation. In the case of the BN/Ag NPs, the fluorescence intensity remained ~90% even after 30 cycles (red) because of the BN spacer's effective passivation of the Ag NPs. This was attributed to the defect-free structure of the BN dielectric spacer layer, consistent with our previous reports.[23,25]

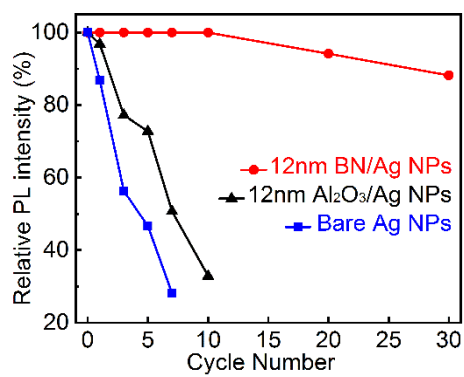

**Figure. 6. Reusability tests. Relative fluorescence intensities for the bare Ag NP (blue), 12 nm $Al_2O_3$/Ag NP (black), and 12 nm BN/Ag NP (red) substrates after different numbers of cycles of heating at 350 °C in the air to remove the R6G molecules for reusability.**

## CONCLUSION

In summary, we studied the use of atomically thin BN as a dielectric spacer to improve the MEF. The small thickness variation of the BN in the atomic range had a large influence on the fluorescence intensity and slightly changed the luminescence wavelength. The 12 nm BN-covered Ag NP substrate exhibited higher EF (up to ~ 95 ± 5), sensitivity ($10^{-8}$ M of R6G), and reusability (only minor decrease in intensity after 30 cycles) than those of the similar MEF substrates covered by traditional spacer materials (*e.g.,* ALD-grown $Al_2O_3$ having the same thickness). According to our PL lifetime measurements and theoretical calculations, both electric field and plasmon enhancement mechanisms contributed to the observed strong MEF. The multifold benefits of the BN spacer will facilitate the applications of MEF in various fields.

## METHODS

**Fabrication of MEF substrates.** All Ag films were deposited at a pressure of ∼ 5 × 10$^{-2}$ mbar and current of 20 mA using a Kaoduen sputtering system. BN and graphene sheets were mechanically exfoliated on the Ag/SiO$_2$/Si substrates. High-quality hexagonal BN single crystals produced by the high-pressure Ba–BN solvent method[52] were used. Suitable sheets were located by optical microscopy (Olympus BX51) and their thicknesses were measured using AFM (Cypher, Asylum Research). The Ag films were heated under an argon (Ar) gas protection at different temperatures to transfer to NPs. A Fiji F200 (Cambridge Nanotech) ALD reactor was used to deposit Al$_2$O$_3$ coatings on Ag NPs, in which trimethyl aluminum (TMA) and water were used as precursors and Ar was used as a carrier and purging gas at a flow rate of 20 sccm. The deposition pressure and temperature were 250 mTorr and 250 °C, respectively. Each deposition cycle consisted of a 0.02 s pulse of TMA, 10 s of Ar purging, 0.06 s pulse of H$_2$O, and 10 s of Ar purging. For the 12 nm Al$_2$O$_3$ coating, 121 cycles were carried out.

**Fluorescence analysis.** An R6G water solution (10$^{-6}$ M) was spin-coated at 2000 rpm for 1 min onto the different MEF substrates using a Laurell spin coater. In the reusability tests, the substrates with R6G were annealed at 350 °C in air for 5 min and re-spin-coated at the same parameters up to 30 times. The PL spectra were acquired using a Renishaw Raman instrument with a 514.5 nm Ar laser and 100× objective lens (numerical aperture: 0.9). The following formula was used to convert wavenumber (cm$^{-1}$) to wavelength (nm):

$$\Delta\omega(\text{cm}^{-1}) = \left(\frac{1}{\lambda_0(\text{nm})} - \frac{1}{\lambda_1(\text{nm})}\right) \times \frac{(10^7 \text{ nm})}{(\text{cm})}, \qquad (6)$$

where $\Delta\omega$ (cm$^{-1}$) is the Raman shift, $\lambda_0$ (nm) is the excitation wavelength, and $\lambda_1$ (nm) is the Raman spectrum wavelength. The PL mapping step size was 0.5 μm. The lifetime measurements were performed with a 512 nm pulsed excitation laser (PoL051X, Advanced Laser Diode Systems GmbH), with a pulse width of 100 ps and repetition rate of 10 MHz.

**Theoretical modeling and calculations.** The finite-element calculations were performed with Comsol Multiphysics 5.1, RF module. Ag NPs described by the dielectric function of Johnson and Christy[53] were placed on the SiO$_2$/Si (the permittivity was assumed to be 5) substrate and covered with a homogeneous spacer (permittivities: 4.5 for BN, 2.56 for Al$_2$O$_3$). The structure was meshed with an average of 100,000 elements, with maximum and minimum sizes of 20 and 0.8 nm, respectively, while 400-nm-thick perfectly matched layers were used to minimize the reflections at the simulation area boundaries. The R6G fluorophores were modelled as point dipoles on top of the BN- and Al$_2$O$_3$-covered Ag NP dimers (Supporting Information, Figure S5). For the excitation, the structures were illuminated by a normally incident plane wave polarized along the dimer axis. The electric fields at the positions of the three dipoles were recorded. For the emission, point dipoles at the same positions were used as radiating sources. The quantum yield was evaluated by calculating the radiated ($P_r$) and absorbe$\Delta$d ($P_{nr}$) powers in a sufficiently large volume around the dimers. The EF was then obtained as the average over the three excitation rates $|\boldsymbol{p_i} \cdot \boldsymbol{E_i}|^2$ (where $\boldsymbol{p_i}$ is the dipole moment of emitter $\boldsymbol{i}$ experiencing a field $\boldsymbol{E_i}$) multiplied by the quantum yield $P_r/(P_r+P_{nr})$.

**ASSOCIATED CONTENT**

**Supporting Information**. Typical AFM images of the exfoliated graphene and BN on 70 nm Ag films and their corresponding PL spectra. SEM images of Ag NPs and corresponding UV-Vis spectra. Schematics of the models for theoretical calculations. Typical PL spectra of R6G, RB on different substrates. This material is available free of charge *via* the Internet at http://pubs. acs.org.

The authors declare no competing financial interest.

## AUTHOR INFORMATION

**Corresponding Author**

luhua.li@deakin.edu.au

**Author Contributions**

L.H.Li conceived and directed the project. W.Gan and Q.Cai fabricated the few-layer hBN and graphene samples; W.Gan did PL measurements. C.Tserkezis performed the theoretical calculations. A.Falin did the AFM measurements. M.Nguyen and I.Aharnonvich measured the lifetime. K.Watanabe and T.Taniguchi provided the single crystal *h*BN. F.Huang and L.Song involved in the discussion. L.X.Kong helped with the ALD deposition. S.Mateti and Y.Chen did the SEM. L.H.Li, W.Gan, and C.Tserkezis wrote the manuscript with inputs from all other authors.

## ACKNOWLEDGMENTS

*ACS Nano* 2019, 13, 12184−12191
DOI: 10.1021/acsnano.9b06858

L.H.Li and W.Gan thank the financial support from the Australian Research Council (ARC) *via* Discovery Early Career Researcher Award (DE160100796). Q.Cai acknowledges ADPRF from Deakin University. C.Tserkezis acknowledges support from VILLUM FONDEN (grant no. 16498), and simulation support from the DeIC National HPC Centre, SDU. K. I.Aharonovich thanks funding from ARC DP180100077 and DP190101058. Watanabe and T. Taniguchi acknowledge support from the Elemental Strategy Initiative conducted by the MEXT, Japan and the CREST (JPMJCR15F3), JST. Part of the work was done at the Melbourne Centre for Nanofabrication (MCN) in the Victorian Node of the Australian National Fabrication Facility (ANFF).